\documentclass[aps,superscriptaddress,preprint,nofootinbib,eqsecnum]{revtex4}
\usepackage{graphicx}
\usepackage{rotating}
\usepackage{latexsym}

\newcommand{\beq}{\begin{equation}}
\newcommand{\eeq}{\end{equation}}
\newcommand{\bqa}{\begin{eqnarray}}
\newcommand{\eqa}{\end{eqnarray}}

\begin{document}

\title{Relativistic (Lattice) 
Boltzmann Equation with Non-Ideal Equation of State}

\author{Paul Romatschke}
\affiliation{
    Frankfurt Institute for Advanced Studies,
    D-60438 Frankfurt, Germany 
}
\affiliation{Department of Physics, 390 UCB, University of Colorado, Boulder,
CO 80309, USA}

\begin{abstract}
The relativistic Boltzmann equation for a single particle
species generally implies a fixed, unchangeable equation of state
that corresponds to that of an ideal gas.
Real-world systems typically have more complicated 
equation of state which cannot be described by the Boltzmann equation.
The present work derives a 'Boltzmann-like' equation that gives
rise to a conserved energy-momentum tensor with an arbitrary 
(but thermodynamically consistent) equation of state. Using
this, a Lattice Boltzmann scheme for diagonal
metric tensors and arbitrary equations of state is constructed.
The scheme is verified for QCD in the Milne metric 
by comparing to viscous fluid dynamics.
\end{abstract}

\date{Late August, 2012}

\maketitle

\section{Motivation}

The Boltzmann equation is a tool that has proven to be very useful in many
different areas of physics. Despite its usefulness, there are some
properties of the Boltzmann equation that are not optimal
for modelling physics systems. 
In particular, for a single particle species
the equation of state is fixed by one 
parameter alone, namely the particle's mass. Since in the limit
of a small particle mean free path the Boltzmann equation 
describes fluid dynamics, this implies that the equation of state
for the fluid hence described is unchangeable and (typically) not realistic.
This is a problem in particular 
for the so-called Lattice Boltzmann Approach to 
fluid dynamics \cite{LBE1, LBE2, LBE3}, 
where the Boltzmann equation serves as a convenient
algorithm for computing the behavior of fluids.
In the non-relativistic context, ways to circumvent this problem are
known, e.g. a modification of the equilibrium distribution function,
modifying only the pressure components or introducing a new
force term \cite{non-id1,non-id2,non-id212,non-id3,non-id4},
but it is not obvious how to generalize those 
to the relativistic case \cite{Mendoza:2009gm, Romatschke:2011hm}.

On the other hand, it is known that particle masses change 
when considering a heat bath: for instance, photons acquire
temperature-dependent masses in a plasma, which leads to a corresponding 
change of the plasma equation of state \cite{Kraemmer:2003gd}. 
In high-temperature Quantum-Chromodynamics (QCD), these medium-dependent
quasiparticles have been successfully used to model the
QCD equation of state \cite{Peshier:1999ww}.
Is it thus possible to write down a 'Boltzmann-like' equation
for a single particle species with a medium-dependent mass
that can reproduce any thermodynamically consistent equation of state?
The objective of the present work is to give an affirmative answer to this
question by means of an explicit construction. 

Note that in the context of quasiparticle and Nambu-Jona-Lasinio
models, essentially
all the relevant parts of the present derivation can be found
\cite{Blattel:1988zz,Bozek:1998dj,Plumari:2010ah,Bluhm:2010qf,Plumari:2011mk}. 
In this sense, the results presented here are not new.
However, as far as I can tell, all published results employ
multiple species of particles, while the results below are for a single species
of a 'virtual' particle, and therefore probably computationally cheaper.
Also, to my knowledge the present work is the first to provide
a concrete example for an algorithm outside equilibrium with
an arbitrary equation of state. The article is structured as follows:
in Sec.~\ref{sec:review}, I give a textbook-style 
review of the Boltzmann equation in curved spaces. In Sec.~\ref{sec:nonid},
a framework for arbitrary non-ideal equations of state is set up
and subsequently 
tested for the case of QCD at high temperature. In Sec.~\ref{sec:LB},
a relativistic lattice Boltzmann scheme for matter with
a non-ideal equation of state in curved spacetime is given,
with the particular example of QCD in a Milne spacetime
that may be of relevance for high energy nuclear collisions.
Finally, I conclude in section \ref{sec:conc}.

\section{Boltzmann Equation in Curved Space: a Review}
\label{sec:review}

This section gives a text-book style review of the Boltzmann equation
in curved space, introducing the usual particle current and 
energy-momentum tensor. Expert readers may want to skip this 
section and read on in Sec.~\ref{sec:nonid}.

The Boltzmann equation specifies the evolution of the single particle
distribution function $f(X^\mu,P^\mu)$, which is dependent on space-time 
$X^\mu\equiv(t,{\bf x})$ and four-momentum $P^\mu\equiv(E,{\bf p})$. 
If collisions are absent, but forces such as gravity are present,
particles are assumed to propagate along geodesics which can be
parameterized by an affine parameter ${\cal T}$. 
Accordingly, the single particle distribution $f$ does not change along
geodesics,
$$
\frac{d f}{d{\cal T}}=\frac{d t}{d{\cal T}}\frac{\partial f}{\partial t}
+\frac{d {\bf x}}{d{\cal T}}\frac{\partial f}{\partial {\bf x}}
+\frac{d P^{\alpha}}{d{\cal T}} \frac{\partial f}{\partial P^{\alpha}}=0\,.
$$
Multiplying with the mass $m$ one can recognize
$m\, dt/d{\cal T}=E$, $m\, d{\bf x}/d{\cal T}={\bf p}$, the 
energy and momentum of a relativistic particle. When re-instating
collisions, particles will no longer follow geodesics,
so $df/d {\cal T}$ will no longer be vanishing. Hence in
the general case one has
\beq
P^\mu \partial_\mu f + F^\alpha \partial_\alpha^{(p)}f = -{\cal C}[f]\,,
\label{Boltzmann}
\eeq
where ${\cal C}[f]$ is the collision term and $F^\alpha\equiv m \frac{d P^{\alpha}}{d{\cal T}}$ the force felt by individual particles.
For gravity, the force is given by $F^\alpha=-\Gamma^\alpha_{\mu\nu}P^\mu P^\nu$
where $\Gamma^\alpha_{\mu\nu}$ are the Christoffel symbols that are calculated
as derivatives of the underlying metric tensor $g_{\mu\nu}$.
For electromagnetism, the force is given by the Lorentz force
$F^\alpha=q F^{\alpha \beta} P_\beta$ where $F^{\alpha \beta}$ is the 
electromagnetic field strength tensor that can be specified in terms
of electric and magnetic fields, and $q$ is the particle's charge.

Including both the gravitational and electromagnetic force terms, 
let us now take an integral moment of Eq.~(\ref{Boltzmann}) with weight
\beq
\label{measure}
\int d\chi \equiv \int \frac{d^4 P}{(2\pi)^4} \sqrt{-g}\, 2\Theta(p^0) (2\pi) \delta\left(
g_{\mu\nu} P^\mu P^\nu-m^2\right)\,,
\eeq
where for clarity $d^4P=\prod_{\mu=0}^3 dP^\mu$ and $\Theta$ denotes the Heaviside step-function, 
$g$ denotes the determinant of the metric tensor $g_{\mu\nu}$
and I have adopted the 'mostly-minus' sign convention for the metric.
The delta-function in $d\chi$ places particles 
on the mass shell and the step-function picks out positive energy states. 
Apart from the appearance of $p^0$, which
could be replaced by a scalar product with a future pointing four-vector,
this form of $d\chi$ is Lorentz covariant (cf.~\cite{GLW}).
Using $\partial_\mu \sqrt{-g}=\sqrt{-g}\Gamma_{\alpha \mu}^\alpha$ 
and $\partial_\lambda g_{\mu\nu}=\Gamma_{\lambda \mu}^\rho g_{\rho \nu}+
\Gamma_{\lambda \nu}^\rho g_{\rho \mu}$ 
one has
\bqa
&
\sqrt{-g}P^\mu \partial_\mu f=\nabla_\mu \left(\sqrt{-g}P^\mu f\right)
-2 \sqrt{-g}P^\mu  \Gamma^\alpha_{\alpha \mu}\,f \,,&\nonumber\\
&\nabla_\mu\left[2\Theta(p^0)\delta\left(P^2-m^2\right)\right]=
2\Theta(p^0) \delta'\left(P^2-m^2\right) 2 P^\alpha P_\beta \Gamma^\beta_{\alpha \mu}\,,&\nonumber
\eqa
where $\nabla_\mu$ denotes the (geometric) covariant derivative. Rewriting
$
 2 P_\beta \delta'\left(P^2-m^2\right) = \partial_{\beta}^{(p)} \delta
\left(P^2-m^2\right)
$
and using partial integration one finds
$$
\int d\chi P^\mu \partial_\mu f = \nabla_\mu \int d\chi P^\mu f
+\int d\chi \Gamma_{\alpha \mu}^\beta P^\alpha P^\mu \partial_\beta^{(p)}f\,.
$$
Also, it is straightforward to show that $\int d\chi F^{\alpha \beta}
P_\beta \partial_{\alpha}^{(p)}f =0$ via partial integration and the 
fact that $F^{\alpha \beta}=-F^{\beta\alpha}$. Hence the Boltzmann
equation implies 
\beq
\nabla_\mu \int d\chi P^\mu f = - \int d\chi {\cal C}[f]\,.
\eeq
Defining the particle number current as $N^\mu\equiv \int d\chi P^\mu f$
one finds that the Boltzmann equation implies the covariant
conservation of particle number, $\nabla_\mu N^\mu =0$,
if $\int d\chi {\cal C}[f]=0$.

Taking the integral moment $\int d\chi P^\nu$ of Eq.~(\ref{Boltzmann}),
one finds
\beq
\nabla_\mu \int d\chi P^\mu P^\nu f - q F^{\nu \beta}\int d\chi P_\beta 
= - \int d\chi P^\nu {\cal C}[f]\,.
\eeq
For uncharged particles ($q=0$), and defining the energy-momentum tensor as 
$T^{\mu\nu}\equiv \int d\chi P^\mu P^\nu f$, the Boltzmann
equation implies covariant conservation of energy and momentum 
if 
\beq
\label{EMTC}
\int d\chi P^\nu {\cal C}[f] = 0\,.
\eeq
I will assume the collision term to fulfill Eq.~(\ref{EMTC}) for the 
remainder of this work. For charged particles, the Boltzmann
equation implies
$$
\nabla_\mu T^{\mu\nu}=q F^{\nu\beta} N_\beta\,,
$$
or the change of energy and momentum being caused by the Lorentz force for 
a current $J_\beta\equiv q N_\beta$. For the remainder of this
work, I will deal with uncharged particles ($q=0$). However,
the generalization to charged particles should be straightforward.

\subsection{Equation of State for Uncharged Boltzmann Gas}
\label{sec:BG}

In equilibrium, the energy-momentum tensor is given by ideal hydrodynamics,
\beq
\label{idtmunu}
T^{\mu\nu}_{\rm eq}=\epsilon U^\mu U^\nu- p \Delta^{\mu\nu}\,,
\eeq
where $U^\mu$ is the fluid velocity obeying $U^2=1$ and
$\Delta^{\mu\nu}\equiv g^{\mu\nu}-U^\mu U^\nu$.
The equilibrium energy density $\epsilon$ and pressure $p$ of the system 
are related by the equation of state.
Since Eq.~(\ref{idtmunu}) must correspond to the particle's
energy-momentum tensor in equilibrium, one has
$$
\epsilon = U_\mu U_\nu T^{\mu\nu}_{\rm eq}=\int d\chi \left(P^\mu U_\mu\right)^2
f_{\rm eq},\quad p = -\frac{\Delta_{\mu\nu}}{3} T^{\mu\nu}_{\rm eq}=
-\frac{1}{3}\int d\chi \left[P^2-\left(P^\mu U_\mu\right)^2\right] f_{\rm eq}\,,
$$
which may be conveniently evaluated by performing a Lorentz boost
to the frame where $P^\mu U_\mu=p^0$ (recall that $d \chi$ is Lorentz
covariant).


Let us now consider a specific equilibrium distribution function
for a system of uncharged particles (cf.~\cite{GLW}),
\beq
\label{feq}        
f_{\rm eq}(X^\alpha,P^\alpha)=Z \times \exp{\left[-\left(\frac{P^\alpha U_\alpha-\mu}{T}\right)\right]}\,,
\eeq
where $Z$ denotes the number of degrees of freedom 
and $\mu$ and $T$ are the chemical potential and temperature,
respectively. In Eq.~(\ref{feq}), $U_\alpha$
is a macroscopic velocity that can be identified with the fluid velocity
in Eq.~(\ref{idtmunu}). 
In this case, $\epsilon, p$ and the number density $n\equiv U_\mu N^\mu$ may be evaluated as\footnote{
Note that this definition of $n$ corresponds to $\partial p/\partial \mu$.
To see this, first go to the local rest frame where $P^2-(P^\alpha U_\alpha)^2={\bf p}^2$ and then
rewrite $\partial f_{\rm eq}/\partial \mu=-\partial_0^{(p)}f_{\rm eq}$. 
Integrate by parts and rewrite $2{\bf p}^2 \delta'(P^2-m^2)=-p^i \partial_i^{(p)}\delta(P^2-m^2)$. Another integration by parts then gives
$\partial p/\partial \mu=U_\mu N^\mu$.}
$$
\epsilon = \frac{Z\, e^{\mu/T} m^2 T}{2\pi^2}\left(3 T K_2\left(\frac{m}{T}\right)+m 
K_1\left(\frac{m}{T}\right)\right)\,,
\quad
p = \frac{Z\, e^{\mu/T} m^2 T^2}{2\pi^2}K_2\left(\frac{m}{T}\right)\,,
\quad 
n = p/T\,,
$$
by using the identity 
$
\int_m^\infty (x^2-m^2)^{n+1/2} e^{-x/T}=(2n+1)!! K_{1+n}(m/T)(m T)^{n+1}
$
for modified Bessel functions $K_\alpha$.
It is straightforward to show that these results obey the 
basic thermodynamic relations
\beq
\label{thermo}
\epsilon+p=s T + \mu n\,,\quad
d\epsilon = T ds + \mu dn\,,
\eeq
where $s$ denotes the entropy density. 
From the equation of state, an interesting quantity to calculate
is the speed of sound squared $c_s^2\equiv dp/d\epsilon$. For illustration,
at $\mu=0$ it can be calculated from the above expressions as
$$
c_s^2(T,\mu=0)=\left(3+\frac{m}{T} \frac{K_2(m/T)}{K_3(m/T)}\right)^{-1}\,,
$$
which increases monotonically with temperature from zero to $1/3$. 
Also, the relation $p=n T$ is the equation of state of an ideal gas.
Clearly, non-ideal
equations of state with a non-monotonic behavior of $c_s$ or $p\neq n T$
are not describable in this framework.

In particular, note that changing the behavior of the equilibrium distribution function 
$f_{\rm eq}$ will not change the relation $p=n T$, 
and hence does not provide the freedom needed to
describe a particular non-ideal equation of state that is
dictated by nature. 

\section{Non-ideal Equations of State}
\label{sec:nonid}

As shown in the preceding section, 
the Boltzmann equation (\ref{Boltzmann}) for a single uncharged
particle species leads to equations of state that depend only
on one parameter, namely the particle's mass. In order to 
describe arbitrary equations of state with a single uncharged
particle species I therefore want to consider temperature
(and density) dependent masses $m\rightarrow M(T,\mu)$, motivated
by the fact that in a plasma at high temperature or density
this approach is physically sound \cite{Kraemmer:2003gd}.
The particles described by the Boltzmann equation should
then be regarded as virtual or 'quasi'-particles, but for 
sufficiently non-ideal equations of state, they will no longer correspond
to any real excitations found in nature. However, the virtue
of introducing this virtual particles will be that no long-range
forces or particle mixtures will be necessary to describe 
the macroscopic system dynamics.

One immediate problem that arises when considering medium-dependent
masses is that thermodynamic consistency is no longer guaranteed.
Specifically, basic thermodynamic relations imply that 
\beq
\epsilon+p = T \left.\frac{\partial p}{\partial T}\right|_\mu + 
\mu \left.\frac{\partial p}{\partial \mu}\right|_T\,,
\label{thermcons}
\eeq
which would be violated when inserting $m\rightarrow M(T,\mu)$ 
in the formulas from Sec.~\ref{sec:BG}. To fix thermodynamic
consistency, I propose the following alternate definition for 
the energy-momentum tensor:
\beq
\label{tmunufull}
T^{\mu\nu}\equiv \int d\chi P^\mu P^\nu f + B(T,\mu)\, g^{\mu\nu}\,,
\eeq
where $B(T,\mu)$ is a function that will be determined
by requiring thermodynamic consistency in equilibrium, cf.~(\ref{thermcons}).
Calculating energy density and pressure from (\ref{tmunufull}),
one finds that $B(T,\mu)$ drops out in $\epsilon+p$ and 
that thermodynamic consistency requires
\beq
\label{cdef}
0 = dB+ \frac{1}{2}\int d\chi f_{\rm eq}\, dM^2\,,
\eeq
where I used $2 {\bf p}^2 \delta'(P^2-M^2) = -p^i \partial_i^{(p)}\delta(P^2-M^2)$ and integration by parts.

Considering the concrete example (\ref{feq}), one has explicitly
\bqa
\label{epsp}
&\epsilon = \frac{Z\, e^{\mu/T} M^2 T^2}{2\pi^2}\left[3K_2\left(\frac{M}{T}\right)+\frac{M}{T} K_1\left(\frac{M}{T}\right) \right]+B(T,\mu)\,,&\nonumber\\
&p  = \frac{Z\, e^{\mu/T} M^2 T^2}{2\pi^2}K_2\left(\frac{M}{T}\right)
-B(T,\mu)\,,\qquad
n  = \frac{Z\, e^{\mu/T} M^2 T}{2\pi^2} K_2\left(\frac{M}{T}\right)\,.&
\eqa

\subsection{Example: QCD at Small Densities}
\label{sec:QCDsmall}

\begin{figure}[t]
\includegraphics[width=.45\linewidth]{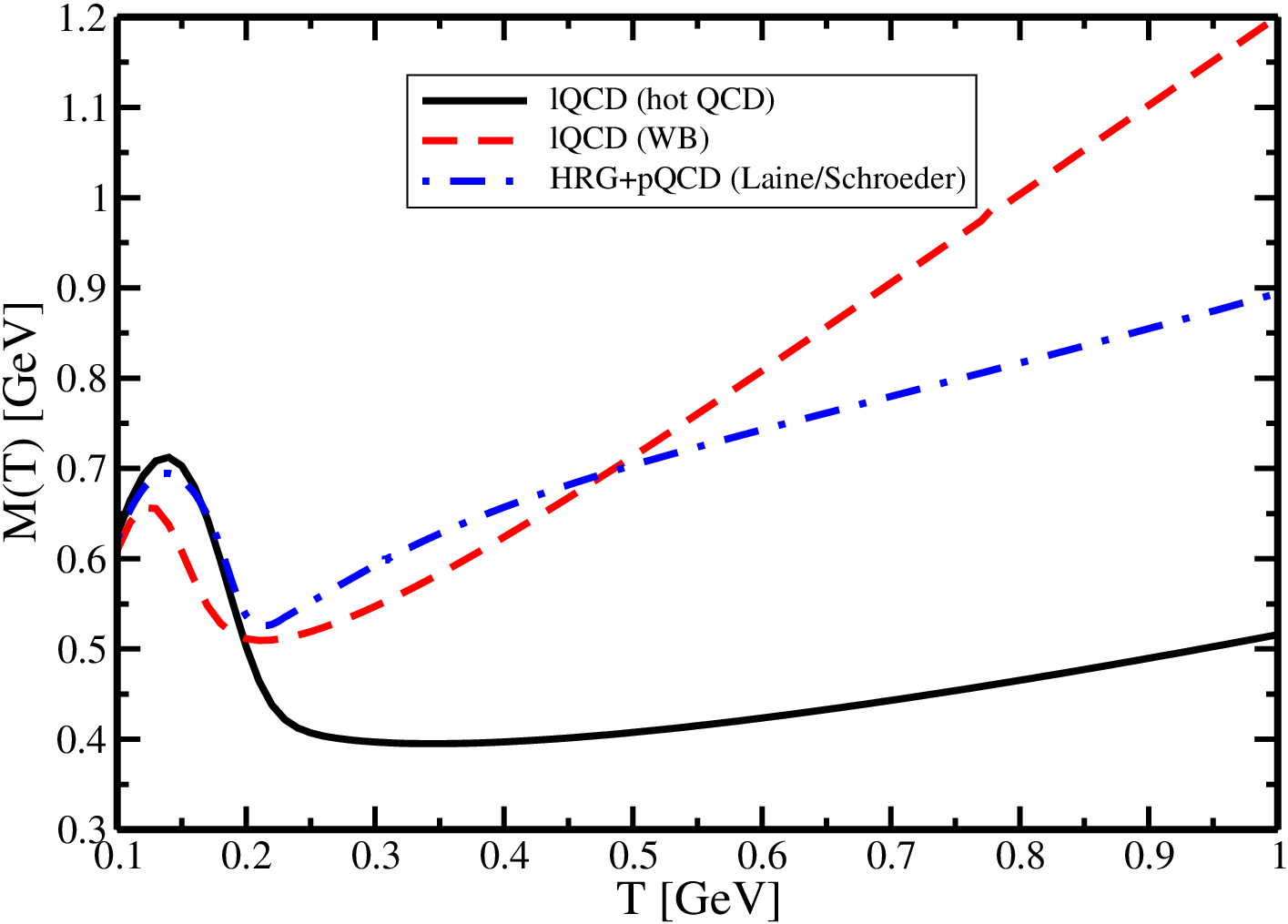}
\hfill
\includegraphics[width=.45\linewidth]{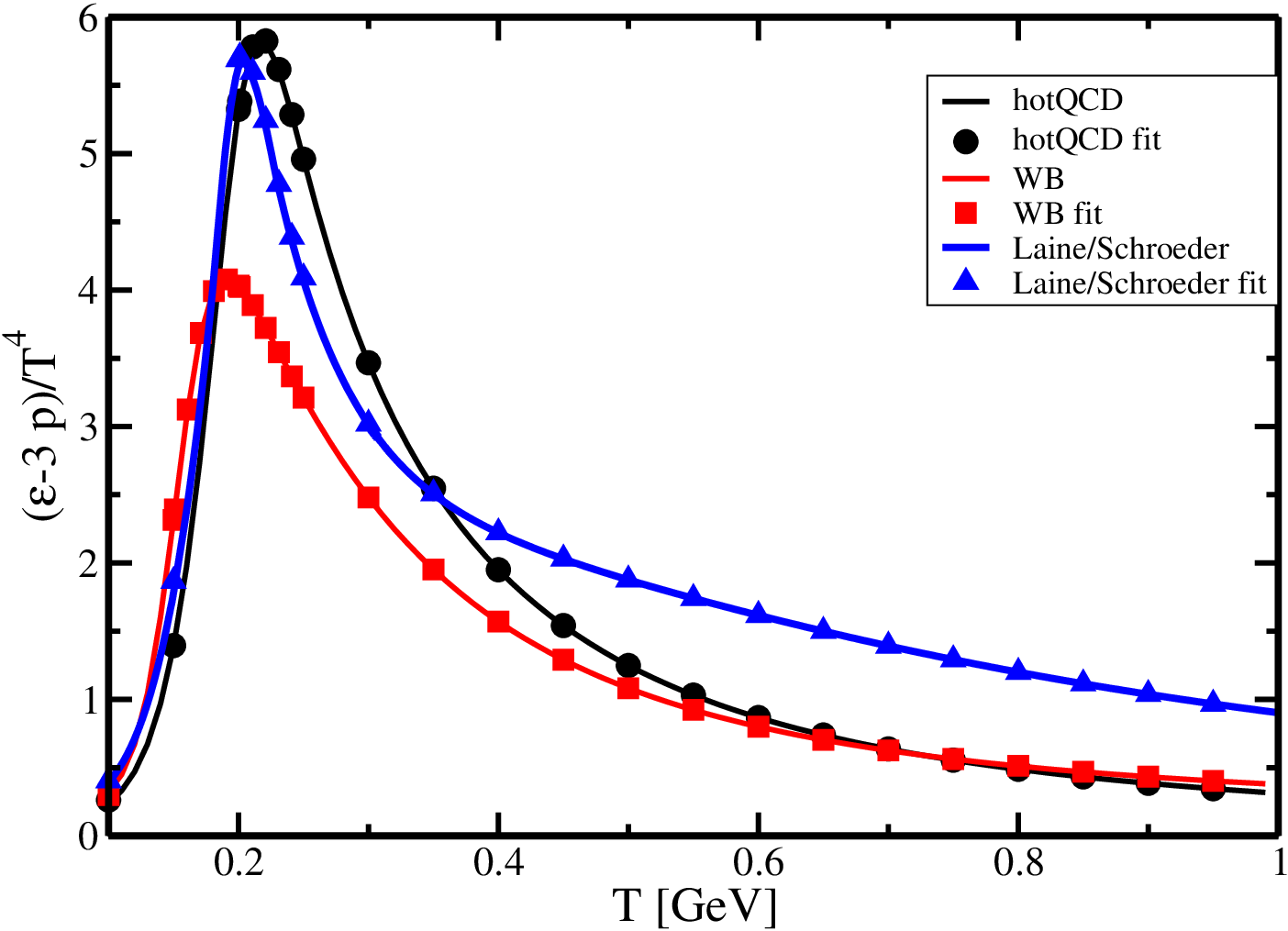}
\caption
{\label{fig:QCDfit}
Left: Results for $M(T)$ when fitting the entropy density
from lattice QCD collaborations 
(hotQCD \cite{Bazavov:2009zn} and Wuppertal-Budapest (WB) 
\cite{Borsanyi:2010cj}, respectively) or an interpolation
from hadron-resonance gas to perturbative QCD (Laine/Schr\"oder,
Ref.~\cite{Laine:2006cp}).
Right: quality of the fit (symbols) when comparing
the trace anomaly $\epsilon-3p$ to the original lattice QCD results
(full lines).}
\end{figure}

Let us consider the above construction for 
the QCD equation of state at zero baryon chemical potential.
In order for the 
Boltzmann energy-momentum tensor to correctly reproduce
the high temperature limit of QCD with $N_c=3,N_f=3$, one has to set
$$
Z = \frac{\pi^4}{180}\left(4 (N_c^2-1)+7 N_c N_f\right)\,.
$$
Then, one can determine $M(T)$ by inverting
$
\frac{\epsilon+p}{T}=\frac{Z}{2\pi^2} M^3 K_3\left(\frac{M}{T}\right)
= s_{lQCD}$, 
where $s_{lQCD}$ may be obtained from the lattice QCD results
(cf.~\cite{Bazavov:2009zn,Borsanyi:2010cj}, both $N_f=3$) or 
an interpolation from hadron resonance gas results to perturbative
QCD (cf.~\cite{Laine:2006cp}, $N_f=4$).
Thermodynamic consistency requires
$
\label{cdef2}
-\frac{Z M^2 T}{2\pi^2} K_1\left(\frac{M}{T}\right) \frac{dM}{dT} = \frac{d B(T)}{dT}\,,
$
which can be solved for $B(T)$ numerically by
integrating up from small temperatures where $B\simeq 0$.
The resulting fits for the masses and the quality of the fit
for the quantity $\epsilon-3 p$ for three 'physical' 
QCD equations of state are shown in Fig.~\ref{fig:QCDfit}.

\subsection{The 'Boltzmann-like' Equation}

The modified energy-momentum tensor (\ref{tmunufull}) is
no longer expected to correspond to a moment of 
the Boltzmann equation (\ref{Boltzmann}), because of the extra term
in (\ref{tmunufull}). However, one can ask if there is a modified
'Boltzmann-like' equation that will give $\nabla_\mu T^{\mu\nu}=0$
(for uncharged particles). Inverting the steps leading to this equation
in Sec.~\ref{sec:review}, 
%
%
and rewriting $P^\mu \delta'(P^2-M^2)=\frac{1}{2}\partial^{\mu}_{(p)}\delta(P^2-M^2)$
and integrating by parts I find that
$$
\nabla_\mu T^{\mu\nu}=\int d\chi P^\nu\left[P^\mu \partial_\mu-\Gamma^\lambda_{\alpha\beta}P^\alpha P^\beta\partial_{\lambda}^{(p)}
+\frac{1}{2}\partial_\mu M^2  \partial^\mu_{(p)}\right]f=0\,,
$$
where the term involving $B(T,\mu)$ cancels if
\beq
\label{sanitycond}
0=dB+\frac{1}{2}\int d\chi f dM^2\,.
\eeq
Note that this is the same as 
the thermodynamic consistency condition (\ref{cdef}),
except that it is promoted to hold also out of equilibrium.
As a consequence, the 'Boltzmann-like' equation
\beq
\label{Blike}
P^\mu \partial_\mu f -\Gamma^\lambda_{\alpha\beta}P^\alpha P^\beta\partial_{\lambda}^{(p)} f
+\frac{1}{2}\partial_\mu M^2  \partial^\mu_{(p)} f =-{\cal C}[f]
\eeq
is guaranteed to reproduce a conserved energy-momentum tensor that 
allows arbitrary equations of state parameterized by medium-dependent
masses $M(T,\mu)$. Note that the 'new' term 
$\frac{1}{2}\partial_\mu M^2  \partial^\mu_{(p)} f$ precisely 
corresponds to the result found when deriving
the Boltzmann equation from quantum field theory (cf.~\cite{Berges:2005md}).
%
%
Finally, repeating the steps in Sec.~\ref{sec:review}, 
one finds that for $\int d\chi {\cal C}[f]=0$, Eq.~(\ref{Blike}) leads to 
\beq
\nabla_\mu N^\mu=0\,,\quad N^\mu\equiv \int d\chi P^\mu f\,,
\eeq
so that the current is formally unchanged when considering
medium-dependent masses. 

\subsection{On-shell-ization}
\label{sec:onshell}

For some applications, it is useful to explicitly perform the
$dp^0$ integral of the Boltzmann-like equation. The reason
is that if one is interested in moments of the Boltzmann equation
with respect to the integral measure $d\chi$, this will allow
one to work with a distribution function $\hat{f}$ 
that then only depends on ${\bf p}$ rather than the four momentum $P^\mu$. 
When discretizing momenta (see below), one thus only has to deal
with three dimensions rather than four.
Note that any factors of $p^0$ that one may have wanted to include
\emph{before} integration will simply turn into multiplicative factors
of 
$E\equiv \sqrt{(-g_{ij}p^i p^j+M^2)/g_{00}}$ 
because of the delta-function
that is part of $d\chi$. Defining
\beq
\hat{f}(X^\mu,{\bf p})\equiv \int dp^0 2p^0 \Theta(p^0) \delta(P^2-M^2)  f(X^\mu,p^0,{\bf p})
\eeq
which is in accordance with \cite{Bernstein} up to a factor of $g_{00}$, 
one can integrate Eq.~(\ref{Blike}) with 
$\int dp^0 2 \Theta(p^0) \delta(P^2-M^2) $,
finding
\beq
\label{onshell}
 \partial_\mu \left(P^\mu \frac{\hat{f}}{E}\right)
- \Gamma_{\alpha\mu}^i \partial_i^{(p)}\left( P^\alpha P^\mu \frac{\hat{f}}{E}\right)
+\frac{1}{2}\partial_i M^2 \partial^i_{(p)}
\frac{\hat{f}}{E}
+2 \Gamma_{\alpha \mu}^\alpha P^\mu \frac{\hat{f}}{E}
=-\frac{1}{E}{\cal C}[\hat f]\,,
\eeq
where $P^\mu$ here is to be understood as on-shell momentum, $P^\mu=(E,{\bf p})$.
In terms of $\hat{f}$, the particle current and energy-momentum tensor
are given as
\beq
\label{NTdisc}
N^\mu = \int \frac{d^3p}{(2\pi)^3} \sqrt{-g} P^\mu \frac{\hat{f}}{E}\,,\quad
T^{\mu\nu}=\int \frac{d^3p}{(2\pi)^3} \sqrt{-g} 
P^\mu P^\nu \frac{\hat{f}}{E}
+B(T,\mu) g^{\mu\nu}\,.
\eeq

As a specific example (cf.~\cite{Bernstein}), consider a metric with a line element of the form $ds^2=dt^2-R(t)^2(d{\bf x}^2)$. Then
$
\Gamma^i_{\alpha \mu}P^\alpha P^\mu = 2 E\, p^i R'/R\,,\quad
\Gamma^\alpha_{\alpha \mu}P^\mu = 3 E\, R'/R
$
and Eq.~(\ref{onshell}) becomes
$$
P^\mu \partial_\mu \hat{f} - \frac{2 p^i  E R^\prime}{R} \partial_i^{(p)}\hat{f} 
+\frac{1}{2}\partial_i M^2 \partial^i_{(p)}\hat{f}
 = -{\cal C}[\hat f]\,.
$$

As another example, consider the line element
$ds^2=d\tau^2-dx^2-dy^2-\tau^2 dY^2$ (Milne metric). Then one has
$\Gamma^i_{\alpha \mu}P^\alpha P^\mu = - 2 E\, p^Y g^{iY} \tau\,,\quad
\Gamma^\alpha_{\alpha \mu}P^\mu = E/\tau$, 
so that one finds
\beq
P^\mu \partial_\mu \hat{f}
- \frac{2 p^Y E}{\tau}\partial_Y^{(p)} \hat f
+\frac{1}{2}\partial_i M^2 \partial^i_{(p)}\hat{f}
= -{\cal C}[\hat f]\,.
\label{Bmilne}
\eeq
\section{Lattice Boltzmann-Equations}
\label{sec:LB}

The main idea behind Lattice Boltzmann equations is to have
a minimum sampling of momentum space given by a discrete set
of $N$ vectors $P^\mu_n$ with $n=0,\ldots N-1$ such that the conservation
equations for the current and energy-momentum tensor are reproduced exactly.
For maximum efficiency, one uses a linear ansatz for the collision term
\beq
\label{bgk}
{\cal C}[\hat f]=\frac{P^\mu U_\mu}{\tau_R} \left(\hat f-\hat f_{\rm eq}\right)
\eeq
with $\tau_R$ the relaxation time. Taking the first two moments of
the Boltzmann equation this leads to the conservation of the current
and energy momentum tensor provided that
\beq
\label{sanity}
U_\mu T^{\mu\nu} = U_\mu T^{\mu\nu}_{\rm eq}=\epsilon(T,\mu) U^\nu\,\quad
U_\mu N^\mu = U_\mu N^{\mu}_{\rm eq}= n(T,\mu)\,,
\eeq
where the equilibrium energy and particle densities are given in 
Eqns.~(\ref{epsp}). The function $B$, which is required to always
match its equilibrium value, is determined from Eq.~(\ref{cdef}).

Before discretizing momentum space on a lattice, it is instructive
to first consider the shear and bulk viscosity coefficients
that the collision term (\ref{bgk}) corresponds to.

\subsection{Chapman-Enskog Expansion}
\label{CE}

In the hydrodynamic (close to equilibrium) limit, the particle
distribution function can be expanded around equilibrium in
powers of space-time gradients, 
$$
f=f_{\rm eq}+f_1+f_2+\ldots\,,
$$
where $f_1$ is of first order in gradients, $f_2$ of second order, and so on.
In the absence of external forces ($\Gamma_{\alpha \beta}^\lambda=0$), 
the Boltzmann Equation (\ref{Blike}) with the collision
term (\ref{bgk}) can then be solved iteratively in powers
of gradients. Specifically, to first order in gradients one finds
$$
f_1 = -\frac{\tau_R}{P\cdot U}\left[
P^\mu \partial_\mu f_{\rm eq}+\frac{1}{2}\partial_\mu M^2 \partial^\mu_{(p)}f_{\rm eq}\right]\,,
$$
which can be evaluated easily using $f_{\rm eq}=Z e^{-P\cdot U/T}$.
Since a small-gradient expansion corresponds to an expansion around
ideal hydrodynamics, we may use the equations of ideal hydrodynamics
to simplify the above equations. Specifically, for a metric signature
of the form $+---$ one has (c.f. \cite{Romatschke:2009kr})
$$ 
D \ln s = -\nabla \cdot U\,,\quad 
D U^\alpha = c_s^2 \nabla^\alpha \ln s\,,
$$
where $c_s$ is the speed of sound and
$D\equiv U^\mu \partial_\mu$ and $\nabla^\alpha\equiv \Delta^{\alpha \beta}\partial_\beta$, $\Delta^{\alpha \beta}=g^{\alpha \beta}-U^\alpha U^\beta$.
Using the thermodynamic relation $\frac{dP}{ds}=c_s^2 T = \frac{dT}{ds} s$ 
and consistently ignoring higher order gradient term corrections,
one finds
$$
f_1=f_{\rm eq}\frac{\tau_R}{P\cdot U} \left[\frac{P^\mu P^\nu}{T} \sigma_{\mu\nu}
+\frac{1}{3 T}\left(P^2
-(1-3c_s^2) (P\cdot U)^2
-3 c_s^2 M T \frac{dM}{dT}
\right)\nabla \cdot U
 \right]
$$
where $\sigma_{\mu\nu}=\nabla_{(\mu}U_{\nu)} -\frac{1}{3}\Delta_{\mu\nu}\nabla \cdot U$. Decomposing the full energy momentum tensor into
$$
T^{\mu\nu}\equiv \int d\chi P^\mu P^\nu f+g^{\mu\nu}B=\epsilon U^\mu U^\nu-P \Delta^{\mu\nu}+\pi^{\mu\nu}+\Delta^{\mu\nu}\Pi\,,
$$
where $\pi^{\mu\nu}=2 \eta \sigma^{\mu\nu}$ and $\Pi=\zeta \nabla \cdot U$,
one identifies the shear and bulk parts of the dissipative tensor with
$$
\pi^{\mu\nu} \equiv T^{<\mu \nu>} = \int d\chi P^{<\mu}P^{\nu>} f\,,\quad
\Pi \equiv \frac{1}{3}\Delta_{\mu \nu} T^{\mu \nu}+P=
\frac{1}{3}
\int d\chi \Delta_{\mu\nu}P^{\mu}P^{\nu} f_1
\,.
$$
Note that $\int d\chi (P\cdot U)^2 f_1=0$ because of Eq.~(\ref{EMTC}).
Using usual the decomposition of the integrals in a tensor basis spanned by
$U^\mu U^\nu$ and $\Delta^{\mu\nu}$ one finds
the shear and bulk viscosity coefficients from $\pi^{\mu\nu}$
and $\Pi$ as
\bqa
\eta&=& \frac{\tau_R}{15 T} Z \int \frac{d^3p}{(2\pi)^3} \frac{(M^2-E^2)^2}{E^2} e^{-E/T}\,,\\
\zeta &=&  \frac{\tau_R}{9 T} Z \int \frac{d^3p}{(2\pi)^3} p^2
\frac{-M^2+(1-3 c_s^2)E^2 +3 c_s^2 M T dM/dT}{E^2} e^{-E/T}\,,
\eqa
where $E=\sqrt{M^2+{\bf p}^2}$. After a little bit of algebra it is possible to show that in the massless limit $\eta=\tau_R \frac{(\epsilon+P)}{5}$
(c.f.~\cite{Boling}), while for constant masses
$\eta = \frac{\tau_R}{T} \int_0^T dT^\prime \left(\epsilon+P\right)$,
$\zeta = \frac{\tau_R}{3T}\left(-3 c_s^2 T (\epsilon+P)+5 \int_0^T dT^\prime \left(\epsilon+P\right)
\right)$.
No simple formulae seem to exist for medium-dependent masses.
Note also that these results differ from 
Ref.~\cite{Betz:2008me,Romatschke:2009im} (and many others using
the Israel-Stewart ansatz)
because non-linearities where not taken account there properly.

Pushing the Chapman-Enskog expansion to second order 
or following Ref.~\cite{York:2008rr} would be 
desirable to extract all the second-order hydrodynamic
transport coefficients \cite{Romatschke:2009kr}. While
this is left for future work, it is possible to
extract the value of the hydrodynamic relaxation times for the 
shear sector, $\tau_\pi$. Identifying the relaxation
time with the coefficient that is multiplying
$-U^\alpha \partial_\alpha (\eta \sigma_{\mu\nu})$ in $\pi_{\mu\nu}$,
one finds after a little algebra $$\tau_\pi=\tau_R$$ from
the derivative of $f_1$. Note that it is therefore possible
to use the same value of $\tau_\pi$ in numerical simulations
using the Lattice Boltzmann with medium-dependent
masses and second-order hydrodynamics.

\subsection{Lattice Boltzmann with Medium-Dependent Masses}

In the following, I will present a 
minimal set of vectors $P^\mu_n$ that is usable for a general relativistic
Boltzmann equation with medium dependent masses, albeit only
for metric tensors that are diagonal. 
The scheme is constructed by noting that the (on-shell)
equilibrium distribution
function for a Boltzmann gas can be expanded as
$$
\hat{f}_{\rm eq}(X^\mu,{\bf p})=e^{\mu/T-E\, u_0/T}\sum_{n=0}^{\infty}
\left(p^i u_i/T\right)^n /n!\,,
$$
and I recall the definition of $E$ given in Sec.~\ref{sec:onshell}.
If the metric is diagonal, one may rescale the space-like momentum 
components such that
$-g_{ij} p^i p^j\rightarrow 
\delta_{ij}\tilde p^i \tilde p^j\equiv |{\bf \tilde p}|^2$. (Note that this rescaling also changes the form of the Boltzmann
equation.)
Now ${\bf \tilde p}/|{\bf \tilde p}|$ is a unit vector 
that may be parameterized by spherical coordinates (angles $\phi,\theta$).
Setting furthermore
$|{\bf \tilde p}|=M \sinh \xi$ (implying $E=M \cosh \xi/\sqrt{g_{00}} $)
one has the parametrization
$$
P^\mu \equiv M \sinh\xi \left({\rm cotanh}\xi/\sqrt{g_{00}}, \sin\theta \cos\phi/\sqrt{-g_{11}},\sin\theta \sin\phi/\sqrt{-g_{22}},\cos\theta /\sqrt{-g_{33}}\right)
$$
for the momentum in terms of the variables $\xi,\theta,\phi$.
Therefore one has
$$
\hat f_{\rm eq}(X^\mu,P^\mu)=e^{-p^0 u_0/T}\sum_{n=0}^{\infty}
\left({\bf v^i}\right)^n\, a^{(n)}(X^\mu)\,,
$$
where ${\bf v}^i\equiv \tilde {\bf p}^i/M \equiv 
\left(\sinh\xi\sin\theta \cos\phi,\sinh\xi\sin\theta \sin\phi,\sinh\xi\cos\theta\right)$
and $a^{(n)}(X^\mu)$ are some coefficients that are only space-time
(but not momentum-) dependent. Powers of the velocities\footnote{
Note that ${\bf v}$ really is equal to velocity times the Lorentz factor.} ${\bf v}$ 
may be be represented using the polynomials 
$P^{(n)}_{i_1\ldots i_n}\left({\bf v}/|{\bf v}|\right)$
that are orthogonal with respect to the angular integral $d\Omega$
(see Ref.~\cite{Romatschke:2011hm} for details).

This then motivates the ansatz for the general distribution function:
\beq
\label{fansatz}
\hat{f}(X^\mu,\xi,\theta,\phi)=e^{-\frac{M^0}{T_0}\cosh \xi}\sum_{n=0}^\infty \sum_{k=0}^\infty P^{(n)}_{i_1\ldots i_n}\left(\frac{{\bf v}}{|{\bf v}|}\right)\, \sinh^n\xi
\, R_{(k)}(\cosh\xi)\,  a^{(nk)}_{i_1,\ldots i_n}(X^\mu)\,,
\eeq
where $T_0,M_0$ are some reference temperature and mass, respectively,
and $R^k$ are polynomials of degree $k$ that will be defined below.
In practice,  the infinite sums above are truncated at some 
finite order. Furthermore, it turns out that for any even $n$,
the $\sinh\xi$ terms may be represented by the sum over polynomials $R_{(k)}$,
so Eq.~(\ref{fansatz}) may be modified such that there is a single
inverse power of $\sinh\xi$ for every $n$ odd. 
%
%
Replacing continuum momenta $P^\mu$ by
a discrete set requires the condition that the integrals in Eqns.~(\ref{NTdisc})
are represented exactly. For the angular integrals, this requirement
is identical to that of massless particles discussed in 
Ref.~\cite{Romatschke:2011hm}. Note that Eqns.~(\ref{NTdisc}) are then 
evaluated for fixed values of $\xi$ rather than fixed
$E,{\bf p}$, implying another change in the Boltzmann equation 
coming from the space-time dependent masses.
For convenience, a concrete example will be given below.

\subsection{Deriving the Momentum Lattice}
\label{sec:lattice}

Let us quickly review the derivation
of the discrete set of momenta: the angles $\phi$
can be found from the requirement that
$$
\int_0^{2\pi} d\phi (\sin\phi)^a (\cos\phi)^{N_\phi-a-1} = \frac{\pi}{N_\phi}\sum_{l}
(\sin\phi_l)^a (\cos\phi_l)^{N_\phi-a-1}\,,
$$
where $a$ is assumed to be a non-negative integer
smaller than $2N_\phi-1$. Namely, the above integrand can be recast
as a Fourier series involving as highest harmonics 
$\cos[(N_\phi-1)\phi]$ and $\sin[(N_\phi-1)\phi]$. 
Exact representation of the integral
as a sum is possible if the angles are chosen as the nodes of 
functions orthogonal to the integrand. For the Fourier series above, 
there are actually two sets of orthogonal functions: $\cos[N_\phi\phi]$ and 
$\sin[N_\phi\phi]$. Choosing $\sin[N_\phi\phi]$, 
the nodes are given by $\phi=\phi_l=\frac{l \pi}{N_\phi}$, $l=0,1,\ldots 2N_
\phi-1$, which fixes the set for $\phi$. For the discrete
set of angles $\theta$, note that the integrands Eq.~(\ref{NTdisc})
only depend on $\theta$ through $P^\mu$, so that $\sin\theta$
always comes with either $\cos\phi$ or $\sin\phi$. Since any odd
power of $\cos\phi,\sin\phi$ integrates to zero, any non-vanishing
contribution must involve $\sin^2\theta=1-\cos^2\theta$. Hence,
it is sufficient to consider only integrands with powers of $\cos\theta$,
which may be recast as a sum:
$$
\int_{-1}^1 d(\cos\theta) \cos^{2N_\theta-1}\theta = \sum_{j}w_j^\theta
\cos^{2N_\theta-1}\theta_j\,,
$$
where the discrete angles $\theta_j$ are given as the roots of the Legendre
polynomial \hbox{$L_{N_\theta}(\cos\theta_j)=0$}, $j=0,1,\ldots N_{\theta}-1$
and the weight factors $w_j^\theta$ are given
as $w_j^\theta=2/\left[(1-\cos^2\theta_j)\left(L_{N_\theta}^\prime(\cos\theta_j)\right)^2\right]$.

Similarly, the integrands in Eq.~(\ref{NTdisc}) then only depend on 
$\cosh\xi$ and $\sin^2\xi$, since any odd power of 
$\sinh\xi$ would have integrated to zero already. 
Therefore, it is sufficient to consider 
integrals of the form
$$
\int_0^\infty d\xi e^{-z_0\cosh\xi}\sinh^2 \xi \cosh^{2N_\xi-1}\xi = \sum_k w_k^\xi(z_0) \cosh^{2N_\xi-1}\left(\xi_k(z_0)\right)\,,\quad z_0\equiv M_0/T_0\,,
$$
and the nodes $\xi_k$ and weights $w_k^\xi$ are calculated 
from the set of polynomials $R_k$ which are orthogonal on $\int d\xi e^{-z_0\cosh\xi}\sinh^2\xi$. Specifically, one finds
\bqa
&R_0(\xi)=1\,,\quad
R_1(\xi)=\cosh\xi - \frac{K_2(z_0)}{K_1(z_0)}\,,&\nonumber\\
&R_2(\xi)=\cosh^2\xi + \frac{6K_1(-z_0^2 K_0^2+(4+z_0^2)K_1^2)}
{z_0 (z_0^3K_0^3+8 z_0^2 K_0^2 K_1+14 z_0 K_0 K_1^2+2(2-z_0^2)K_1^3-z_0^3 K_2^3} 
R_1(\xi) -\frac{3 K_2+ z_0 K_1}{z_0 K_1}\,,\quad {\rm etc.}&\nonumber
\eqa
Hence the discrete values $\xi_k$ are calculated from 
$R_{N_\xi}(\xi_k)=0$ and the weights $w_k^\xi$ fulfill
$$
\sum_{k=0}^{N_\xi-1} w_k^\xi R_0(\xi_k) = \frac{K_1(z_0)}{z_0}\,,\quad
\sum_{k=0}^{N_\xi-1} w_k^\xi R_m(\xi_k) = 0,\,\quad m=1,\ldots N_\xi-1\,.\nonumber\\
$$
Thus, one finds the following representation of the momentum integrals:
$$
\int \frac{d\Omega}{4\pi}\int_0^\infty d\xi \sinh^2\xi\, 
\hat f(\xi,\theta,\phi)=\sum_{k=0}^{N_\xi-1}\sum_{j=0}^{N_\theta-1}\sum_{l=0}^{2N_\phi-1}
w_{kj}\, \hat f(\xi_k,\theta_j,\phi_l)=\sum_n w_n \hat f(P_n^\mu)\,,
$$
with the weights
$w_{kj}=e^{z_0 \cosh \xi_k} w_k^\xi w_j^\theta/(4 N_\phi)\equiv w_n\,,
$
and where the collective index $n$ runs over all discrete momenta $P_n^\mu$
constructed from the ensemble $\phi_l,\theta_j,\xi_k$.

\subsection{Lattice Boltzmann Algorithm for Milne Spacetime}

In this subsection I give a detailed construction 
of a lattice Boltzmann algorithm with non-ideal QCD equation
of state in an expanding spacetime with
$ds^2=d\tau^2-dx^2-dy^2-\tau^2 dY^2$ (Milne). For simplicity,
I will limit myself to neglecting space dependencies,
which are algorithmically easy to program (cf.~\cite{Romatschke:2011hm}
for a practical example).

Starting with the Boltzmann-equation (\ref{Bmilne})
for the on-shell distribution function, let us first rescale
momenta $p^Y=\tilde p^Y/\tau$ so that $E=\delta_{ij}\tilde p^i \tilde p^j$.
Next, replacing $\tilde {\bf p}^i = M(\tau) {\bf v}^i$ and using Eq.~(\ref{bgk}), Eq.~(\ref{Bmilne}) becomes
\beq
\label{newBeq}
\left.\partial_\tau \hat f\right|_{\bf v} -
\frac{\tilde p^Y}{\tau}\partial_Y^{(\tilde p)} \hat f -\partial_\tau \ln M\ 
{\bf \tilde p}\cdot {\bf \partial}_{(\tilde p)} \hat f = - \frac{\hat f-\hat f_{\rm eq}}{\tau_R}\,.
\eeq
Neglecting space dependencies, one has cylindrical 
symmetry and hence $\hat f =\hat f(\tau, \xi, \theta)$. Therefore
the ansatz for the general distribution function can be simpler than (\ref{fansatz}), namely
\beq
\label{newfansatz}
\hat{f}(X^\mu,\xi,\theta,\phi)=e^{-M^0/T_0\cosh \xi}\sum_{n=0}^\infty \sum_{k=0}^\infty L_{n}\left(\cos \theta \right)\, R_{(k)}(\cosh\xi)\,  a^{(nk)}(\tau)\,.
\eeq
Using the nodes and weights from Sec.~\ref{sec:lattice}, one immediately finds
$$
a^{(ml)}=\frac{(2m+1)}{I(l)}\int \frac{d\Omega}{4\pi}\int d\xi
\sinh\xi^2 L_m(\cos \theta) R_l(\cosh\xi) \hat f = \frac{(2m+1)}{I(l)} \sum_n
w_n \left.L_m R_l \hat f \right|_{P^\mu_n}\,,
$$
where $I(l)=\int d\xi \sinh^2\xi R_l^2(\cosh \xi)$. 

I will not consider conserved particle number, so the only
quantity of interest is 
\beq
\label{emtnew}
T^{\mu\nu}= \frac{M^2}{2\pi^2} \sum_n w_n \hat f(P^\mu_n)
P^\mu_n P^\nu_n +B(T) g^{\mu\nu}\,.
\eeq
Since all spatial dependencies have been neglected, the fluid
velocity is trivial, $U^\mu=(1,{\bf 0})$, and the equilibrium energy 
density is given by the $00$ component of Eq.~(\ref{emtnew}).

A change in the energy density can be calculated via the Boltzmann
equation (\ref{newBeq}):
\bqa
\label{edchange}
\partial_\tau \epsilon
&=&\frac{M^4}{2\pi^2} \sum_n w_n  \cosh\xi_n^2 \left[
\frac{\tilde p^Y}{\tau}\partial_Y^{(\tilde p)} \hat f- \frac{\hat f-\hat f_{\rm eq}}{\tau_R}\right]\,,
\eqa
where I used partial integration and the identity (\ref{sanitycond}),
which becomes
\beq
\label{masschange}
\frac{\partial_\tau M^4}{8\pi^2} \sum_n w_n \hat f(P^\mu_n) 
+\partial_\tau B(T)=0\,.
\eeq
More specifically, a lattice Boltzmann algorithm may hence 
be constructed as follows: a valid initial condition at time $\tau$
consists of specifying $\hat f=\hat f_{\rm cur}$ and an initial temperature
and particle mass, $T_{\rm cur},M_{\rm cur}$. Then, make a prediction
of the change in distribution function and (logarithm of) mass:
\bqa
\delta f_{\rm pred}&=&\left(\frac{\hat S_1}{\tau}+
\frac{\delta \ln M_{\rm pred}}{\delta \tau} \hat S_2-\frac{\hat f_{\rm cur}-\hat f_{\rm eq}(T_{\rm cur})}{\tau_R(T_{\rm cur})}\right)\delta \tau \,,\nonumber\\
\delta \ln M_{\rm pred}&=&- \left.\frac{dB}{d\epsilon}\right|_{T_{\rm cur}}
\frac{\sum w_n \left(\hat{S}_1/\tau-\frac{\hat f_{\rm cur}-\hat f_{\rm eq}(T_{\rm cur})}{\tau_R(T_{\rm cur})}\right)}{\sum w_n \hat f_{\rm cur}}\delta \tau
\eqa
where $\hat S_1$ and $\hat S_2$ are representations of the momentum derivatives 
$\tilde p^Y \partial_Y^{(\tilde p)}$ and $\tilde {\bf p}\cdot \partial_{(\tilde p)} \hat f$ in the form of (\ref{newfansatz})
with coefficients
\bqa
s^{ml}_1&=&-\frac{2m+1}{I(l)}\sum_n w_n \hat f_{\rm cur}\left[
\left(P_m R^{\prime}_l \cosh\xi -R_l\right) \cos^2\theta \tanh\xi^2
+R_l\left(P_m+P_m^\prime (\cos\theta-\cos^3\theta)\right)\right]\nonumber\\
s^{ml}_2&=&-\frac{2m+1}{I(l)}\sum_n w_n 
\hat f_{1}\left[P_m R_l \left(2+\frac{1}{\cosh^2\xi}\right)+
P_mR_l^\prime \cosh\xi \tanh^2\xi\right]\,,\nonumber
\eqa
respectively. Via Eq.~(\ref{edchange}), this leads to a prediction for the 
new temperature $T_{\rm pred}$. These predictions are then corrected using
the trapezoid integration formula
$$
\delta f_{\rm corr}=\frac{\delta \tau}{2} \left(
\left.\partial_\tau \hat f\right|_{\tau}+
\left.\partial_\tau \hat f\right|_{\tau+\delta \tau}
\right)+{\cal O}(\delta \tau)^3\,,\quad \delta \ln M_{\rm corr}=
\frac{\delta \tau}{2} \left(
\left.\partial_\tau \ln M\right|_{\tau}+
\left.\partial_\tau \ln M\right|_{\tau+\delta \tau}
\right)+{\cal O}(\delta \tau)^3\,,
$$
where the values at time $\tau+\delta \tau$ are calculated using
$\delta f_{\rm pred}$ and $\delta \ln M_{\rm pred}$. 
Note that the resulting mass $M_{\rm corr}$ does not necessarily
correspond to the equilibrium particle mass $M(T_{\rm new})$.
As a consequence, I use $f_{\rm eq}=Z e^{-\sqrt{M_{\rm eq}^2+M_{\rm corr}^2\sinh^2\xi}/T}$
for the equilibrium distribution function in the algorithm.
The above steps may be repeated to solve the Boltzmann equation
(\ref{newBeq}) for arbitrary times. The resulting
algorithm leads to time integrated quantities that are accurate
to ${\cal O}(\delta \tau)^2$ (cf.~\cite{Dellar}).

\subsection{Results for Milne Spacetime}

In this section I provide tests of the above Lattice Boltzmann 
algorithm by comparing results to viscous fluid dynamics
for the QCD equation of state of Ref.~\cite{Laine:2006cp}
and a Milne metric. The fluid dynamics equations for the
energy density and quantity $\Phi\equiv T^Y_Y-p$
fulfill the coupled equations \cite{Romatschke:2009im} 
\bqa
\label{hydroeq}
\partial_\tau \epsilon = -\frac{\epsilon+P}{\tau}+\frac{\Phi}{\tau}\,,\quad
\partial_\tau \Phi = -\frac{\Phi}{\tau_\pi}+\frac{4 \eta}{3 \tau_\pi
  \tau}-\frac{4\Phi}{3 \tau}-\frac{\lambda_1}{2 \tau_\pi
  \eta^2}\Phi^2\,, \eqa where $\tau_\pi$ is the relaxation time and
$\lambda_1$ is a self-coupling parameter. While I found 
$\tau_\pi=\tau_R$ in Sec.~\ref{CE}, $\lambda_1$ is currently not 
known. For simplicity, for the hydrodynamic
calculation I will use the values 
$\tau_\pi=5 \frac{\eta}{\epsilon+P}$ and $\lambda_1=\frac{5}{7}\eta \tau_\pi$
that are reported for the massless gas case \cite{Boling}. 

One should keep in mind that --- since the correct
values for $\tau_\pi,\lambda_1$ will differ from this choice in view 
of the findings in section \ref{CE} --- this implies that
the hydrodynamic and Lattice Boltzmann results will not agree in
practice. Note, however, that there is another issue that
prevents perfect agreement between (second-order) hydrodynamics
and Lattice Boltzmann theory even in principle: the reason
is that, even if one were to use the same second-order 
transport coefficients in a numerical second-order hydrodynamics
and a Lattice Boltzmann solver, the two would still
disagree because of the different \emph{third} order gradient
terms. However, for all practical purposes when hydrodynamics
itself can be considered applicable, the difference between
the two numerical schemes could be considered small.

For the Lattice-Boltzmann framework, the QCD
equation of state is parameterized as in Sec.~\ref{sec:QCDsmall}
with a reference value $M_0/T_0=1$ for (\ref{fansatz}). Note that
this reference value 
corresponds to a reference temperature of $T\sim 0.82$ GeV,
This means that results will be most accurate for this temperature 
(cf.~\cite{Romatschke:2011hm}),
and in particular will break down if applied to problems involving
fluid cells with temperatures exceeding two times this reference temperature.
The results at temperatures different than this reference value 
can be improved by increasing the value of $N_{\xi}$, 
but in practice I find that $N_{\xi}\geq 3$ gives adequate results.
\begin{figure}[t]
\includegraphics[width=.45\linewidth]{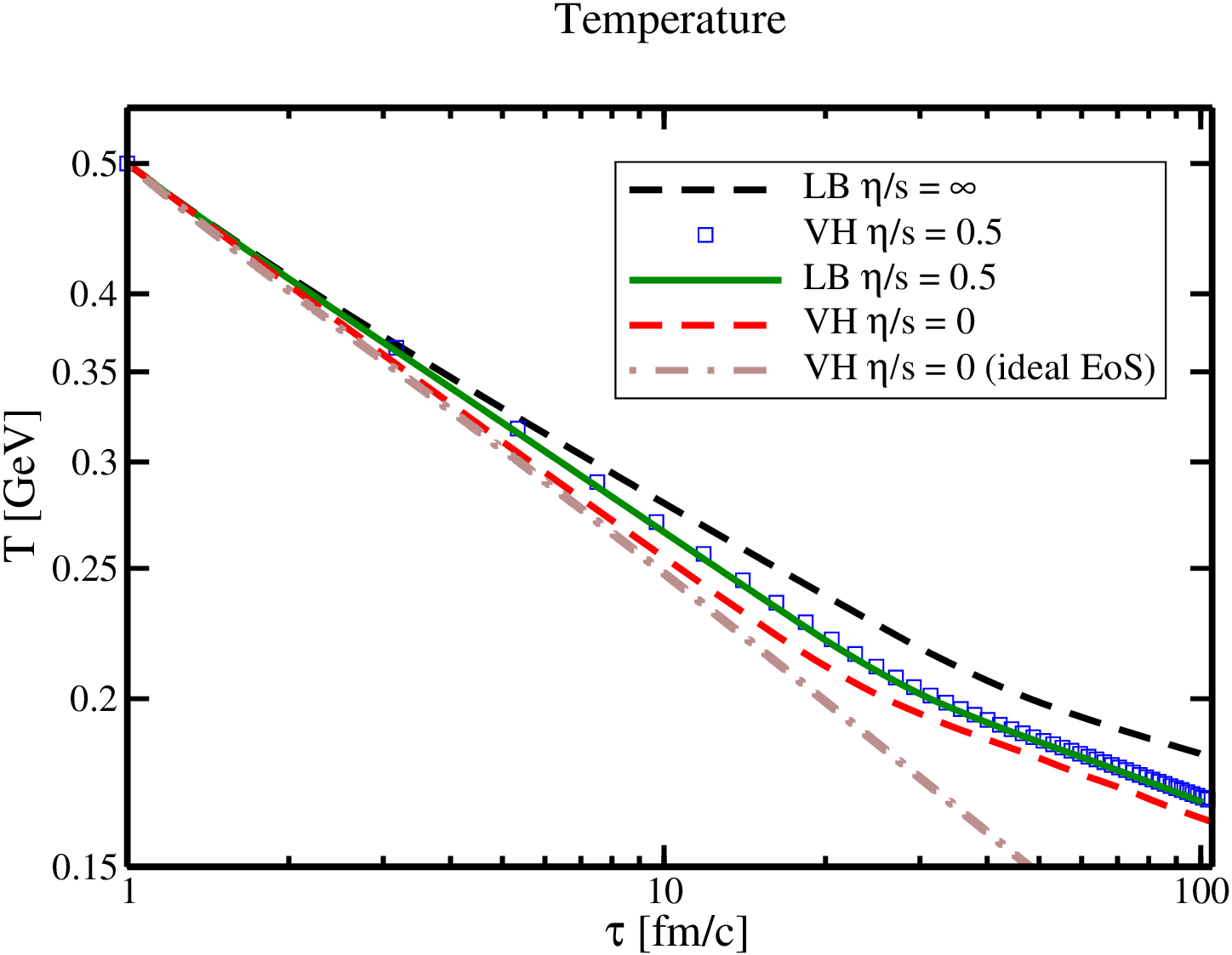}
\hfill
\includegraphics[width=.45\linewidth]{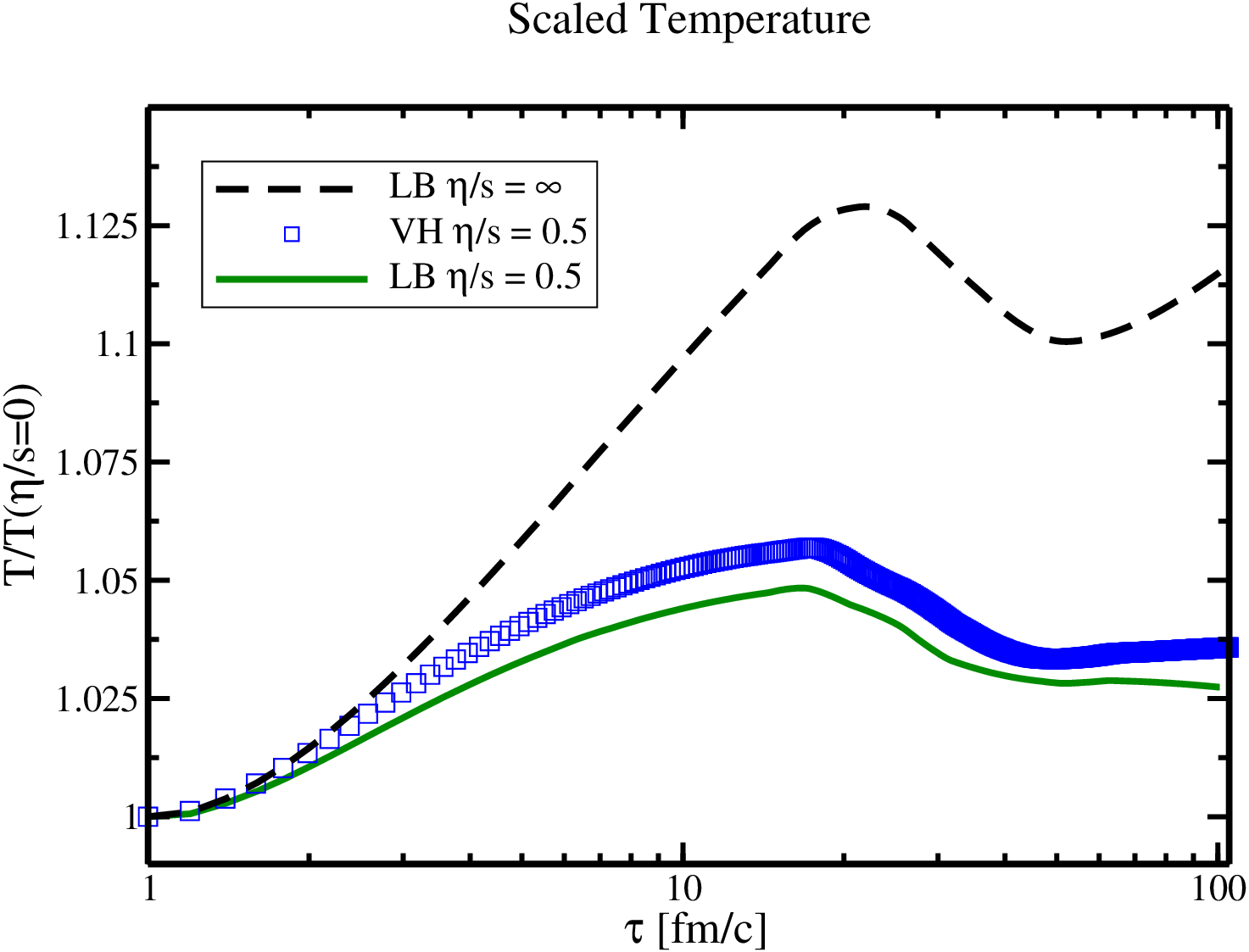}
\caption
{\label{fig:comp}
Temperature evolution for various viscosities,
from fluid dynamics ('VH') and Lattice Boltzmann ('LB') respectively.
Results are for a QCD equation of state except for the ideal
equation of state evolution (left plot).
Transport coefficients for VH are only approximate, so perfect 
agreement is not expected.
Right: results normalized with respect to the ideal fluid dynamics
to highlight differences. The $\eta/s=\infty$ results are 
obtained by setting $\tau_R=\infty$ (free-streaming)
and updating the medium-dependent mass according to the change in
energy density. These results do not correspond to an actual physics
situation and are presented for illustrative purposes only.}
\end{figure}
In Fig.~\ref{fig:comp}, I show the temperature evolution in viscous fluid
dynamics and the above lattice Boltzmann algorithm for $N_\xi=5,N_\theta=5$
and $\delta \tau=0.2$ fm/c. As can
be seen from this figure, the non-ideal equation of state time
evolution in fluid dynamics for $\eta/s=0.5$ is described reasonably well
throughout the whole simulation time, even though it differs markedly
from the ideal equation of state time evolution (shown in Fig.~\ref{fig:comp}
for comparison). 
Overall, the algorithm seems to perform rather well and provides
a concrete example for simulating the Boltzmann equation for
a system with non-ideal equation of state and non-Minkowski 
geometry.

\section{Conclusions}
\label{sec:conc}

In this note I have set up a general relativistic transport equation 
for a single species of uncharged particles with a medium-dependent mass.
This 'Boltzmann-like' equation allows for a conserved particle current
and energy-momentum tensor. The latter explicitly allows arbitrary
(thermodynamically consistent) equations of state when using
the medium-dependent mass as fitting parameter. I expect this
formulation to be useful for relativistic 
fluid dynamics simulation in the Lattice Boltzmann framework 
with arbitrary equations of state. Possibly, it can also have relevance
for non-relativistic computational non-ideal fluid dynamics 
(cf.~\cite{Succi:2008}). Furthermore, it can be applicable 
in the context of simulating parton
dynamics with non-ideal equations of state, 
cf.~\cite{Xu:2008av,Xu:2010cq} or quasiparticle models of QCD
\cite{Bluhm:2010qf}.

\section*{Acknowledgements}
I would like to thank I.~Bouras, A.~El, V.~Greco, M.~Mendoza, D.~Radice
and S.~Succi for fruitful discussions. 
This work was supported in part by the Helmholtz International
Center for FAIR within the framework of the LOEWE program launched
by the state of Hesse and in part by the US Department of Energy 
within the framework of the JET Collaboration under grant No.
DE-AC-02-05CH11231, subcontract No. 6990498. 
This sponsorship does not constitute
endorsement by the University or Governement of the views expressed
in this publication.

\end{document}